\documentstyle[aps,prl,epsfig,twocolumn,floats]{revtex}

\begin{document}
\draft
\twocolumn[\hsize\textwidth\columnwidth\hsize\csname@twocolumnfalse%
\endcsname
\title{Segregation of granular binary mixtures by a ratchet mechanism}
\author{Z\'en\'o Farkas$^{1,2}$,
Ferenc Szalai$^{1,3}$,
Dietrich E. Wolf$^2$, and Tam\'as Vicsek$^1$
}
\address{$^1$Department of Biological Physics, E\"otv\"os University,
Budapest, P\'azm\'any P. Stny 1A, 1117 Hungary\\
$^2$Department of Theoretical Physics, Gerhard-Mercator University, 
D-47048 Duisburg, Germany\\
$^3$ Computer and Automation Research Institute, Hungarian Academy of
Sciences, Budapest, P.O.B. 63, 1518 Hungary
}
\date{13 February, 2001}

\maketitle

\begin{abstract}
We report on a segregation scheme for granular binary mixtures,
where the segregation is performed by a ratchet mechanism
realized by a vertically shaken asymmetric sawtooth-shaped
base in a quasi-two-dimensional box.
We have studied this system by computer simulations
and found that most binary mixtures can be segregated
using an appropriately chosen ratchet, even when the particles
in the two components have the same size, and
differ only in their normal restitution coefficient or
friction coefficient.
These results suggest that the components  of otherwise non-segregating
granular mixtures may be separated using our method.
\end{abstract}

]

%
%
%

\newcommand\nc[2]{\newcommand#1{#2}}
\nc{\rnc[2]}{\renewcommand#1{#2}}
\nc{\ds}{\displaystyle}
\nc{\scs}{\scriptsize}
\nc{\be}{\begin{equation}}
\nc{\ee}{\end{equation}}
\nc{\bea}{\begin{eqnarray}}
\nc{\eea}{\end{eqnarray}}
\nc{\f[2]}{\frac{#1}{#2}}
\nc{\mb[1]}{\mbox{#1}}
\nc{\idx[1]}{{\mb{\scs #1}}}

\nc{\vibrampl}{A}
\nc{\vibrfreq}{f}
\nc{\velunit}{\,\,\frac{\mb{\scs cm}}{\mb{\scs s}}}
\nc{\diffunit}{\,\,\frac{\mb{\scs cm}^2}{\mb{\scs s}}}
\nc{\boxwidth}{L}
\nc{\loadrate}{R}
\nc{\loadheight}{H}
\nc{\vel}{v}
\nc{\probdens}{W}
\nc{\current}{j}
\nc{\leftside}{\idx{$\leftarrow$}}
\nc{\rightside}{\idx{$\rightarrow$}}
\nc{\diffcoeff}{D}
\nc{\ratio}{\alpha}
\nc{\lengthunit}{{l^*}}
\nc{\timeunit}{{t^*}}
\nc{\dt}{\partial_t}
\nc{\dx}{\partial_x}
\nc{\dxx}{\partial^2_x}
\nc{\tilt}{\theta}
\nc{\avrg[1]}{\langle #1 \rangle}
\nc{\dphi}{\delta\phi}

\nc{\Hz}{\,\mb{s$^{-1}$}}
\nc{\mm}{\,\mb{mm}}
\nc{\cm}{\,\mb{cm}}
\nc{\cmps}{\,\mb{cm}\,\mb{s}^{-1}}
\nc{\cmsps}{\,\mb{cm$^2$}\,\mb{s}^{-1}}

%
%

\newcommand\figsetupfile{\centerline{\epsfig{figure=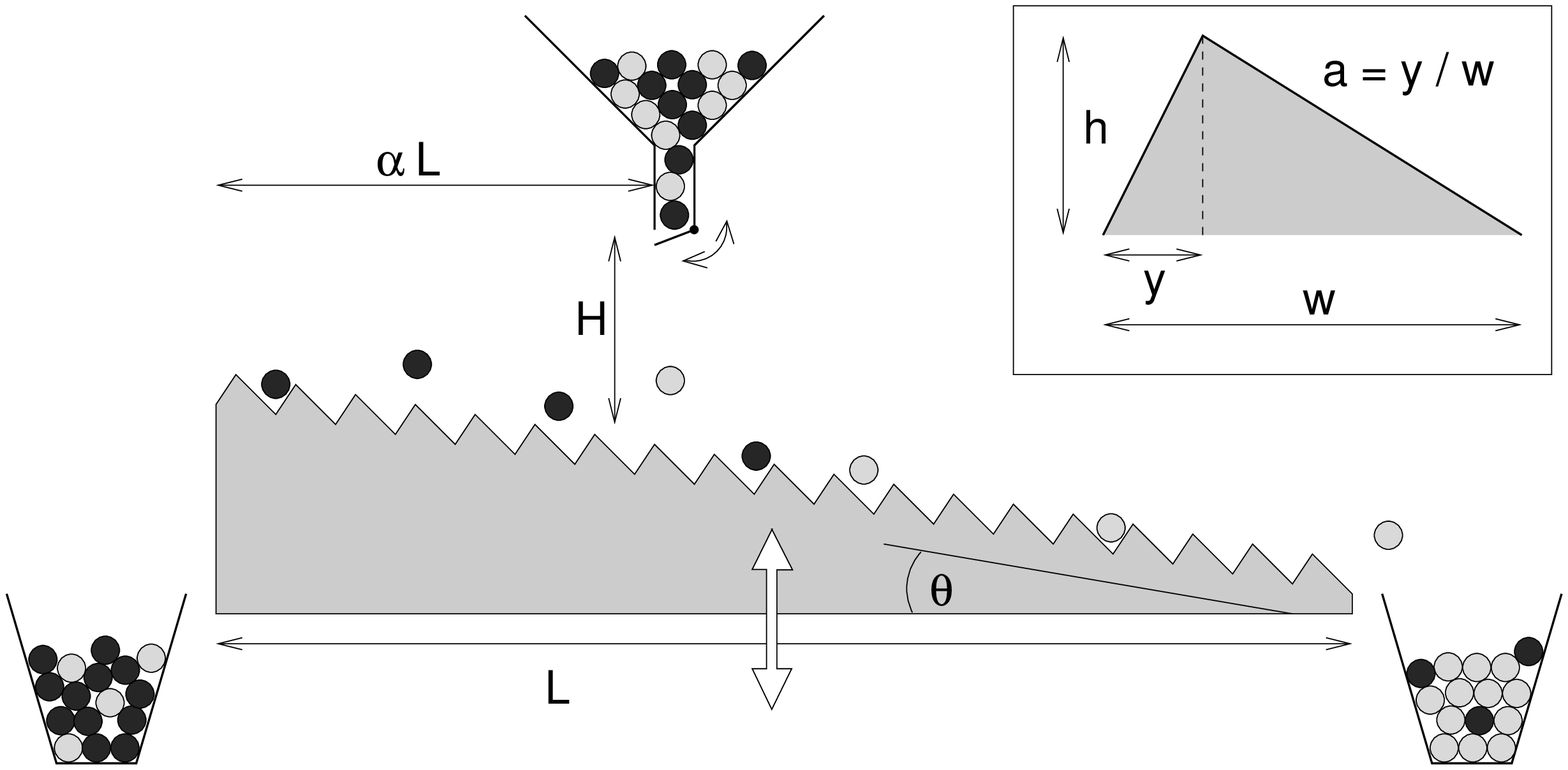,width=\linewidth}}}
\newcommand\figsetup[1]{
%
%
\begin{figure}
#1
\caption{Schematic drawing of the segregating setup.
The particles are falling into the box of width $L$ at
$\ratio \boxwidth$ from the left side of the box ($0 < \ratio < 1$)
from height $H$ with load rate $\loadrate$.
The base is shaken sinusoidally with amplitude $A$ and frequency $f$,
and may be tilted by angle $\tilt$, which is positive if the
left end is higher than the right one.
The particles within a mixture component are exactly the same if not
stated otherwise.
{\em Inset:}
The shape of one sawtooth is described by three parameters: 
width $w$, height $h$ and asymmetry parameter $a$.}
\label{fig:setup}
\end{figure}
}

\newcommand\figdriftdifffile{\centerline{\epsfig{file=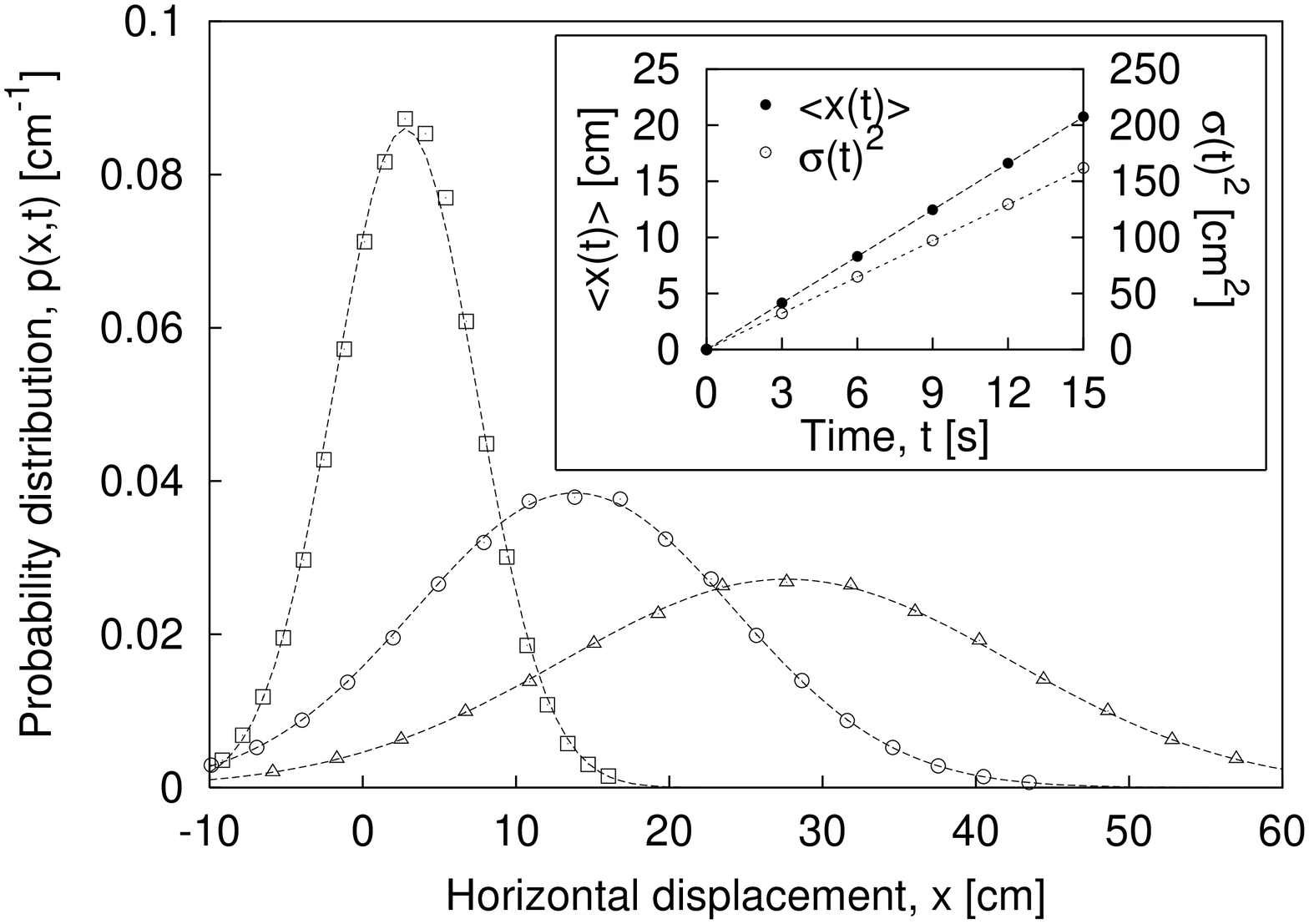,height=\linewidth,angle=270}}}
\newcommand\figdriftdiff[1]{
%
%
\begin{figure}
#1
\caption{Drift-diffusion of a single particle
in an infinitely wide system
(realized by periodic boundary condition).
The probability distribution
of the horizontal displacement $x$ in
time $t=2$ s (square), $t=10$ s (circle) and $t=20$ s (triangle).
The dashed curves show the corresponding
Gaussian distributions with center $v t$ and
dispersion $\sqrt{2 D t}$
as theoretical prediction \protect\cite{risken89}.
The drift velocity $v$ and diffusion constant $D$ are determined
by line fitting:
$\avrg{x(t)} = v t$ and $\sigma(t)^2 = 2 D t$
(see {\em inset}), where
$\sigma(t)^2 = \avrg{x(t)^2} - \avrg{x(t)}^2$.
The fitted values are $v=1.38 \cmps$ and $D=5.38 \cmsps$.
The particle parameters are
$r = 1.5 \mm$, $e = 0.6$, $\mu = 0.3$ and $\beta_0 = 0.4$,
the ratchet parameters are
$w = 12 \mm$, $h = 8 \mm$, $a = 0.2$,
$\theta = 0$, $A = 2 \mm$ and $f = 20 \Hz$.}
\label{fig:driftdiff}
\end{figure} 
}

\newcommand\figsegrfile{\centerline{\epsfig{file=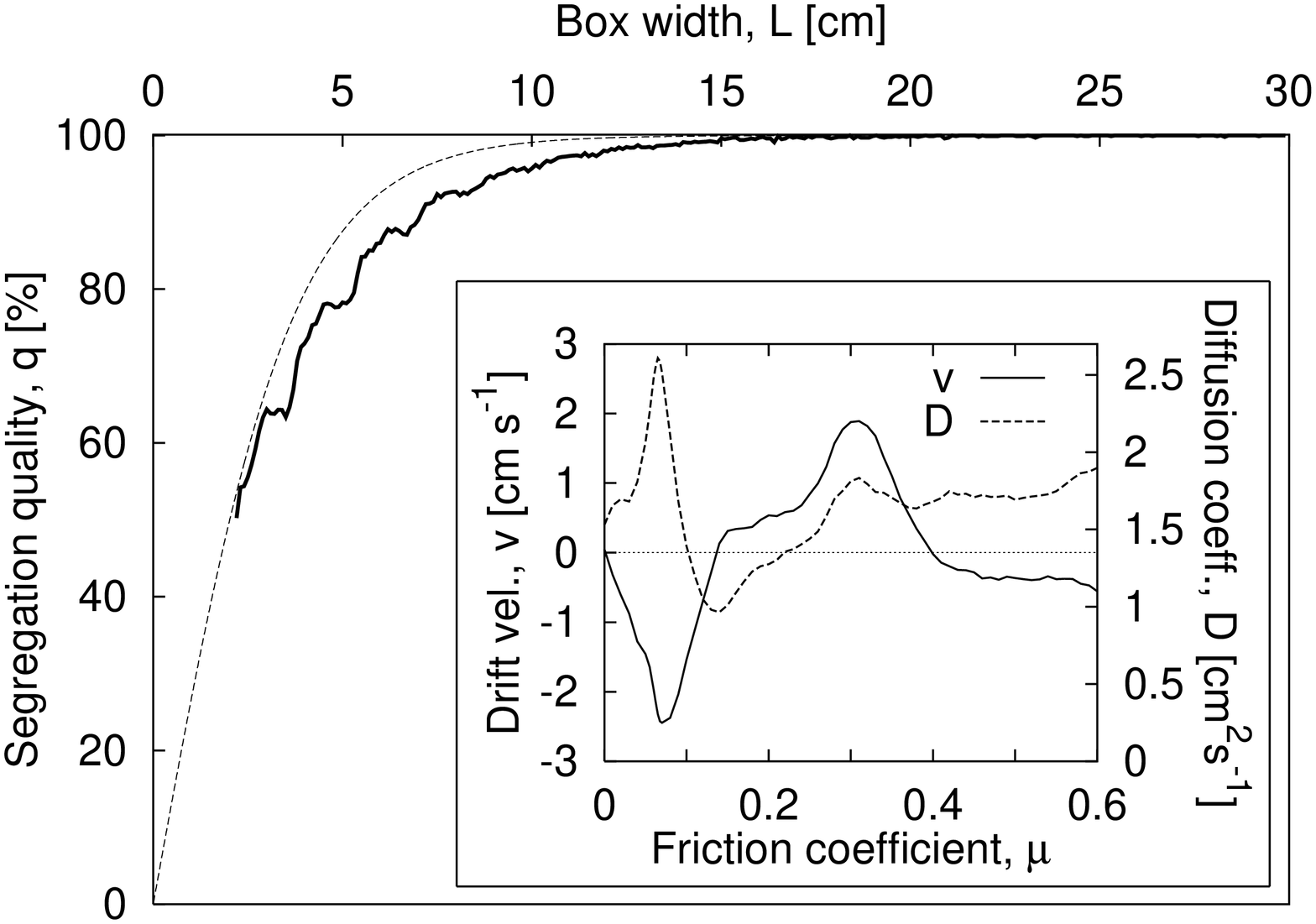,height=0.95\linewidth,angle=270}}}
\newcommand\figsegr[1]{
%
%
\begin{figure}
#1
\caption{Segregation of particles differing only in friction coefficient.
The segregation quality (solid line)
rapidly grows to 100\% with
increasing box width, the dashed line shows the theoretical prediction.
At small box widths the deviation from the theoretical prediction
(dashed line) is due to that
(1) the drift-diffusion approximation is valid
on length scales larger than the sawtooth width, and
(2) the starting state (position and velocity)
of a particle is untypical to the drift-diffusion motion.
{\it Inset:} The drift velocity and diffusion coefficient
(measured in an infinitely wide box)
as functions of the friction coefficient $\mu$.
The drift velocity changes its sign twice, at $\mu = 0.14$ and
at $\mu = 0.4$,
and is only coincidentally zero at $\mu = 0$.
}
\label{fig:segr}
\end{figure} 
}

\newcommand\figsegrallfile{\centerline{\epsfig{file=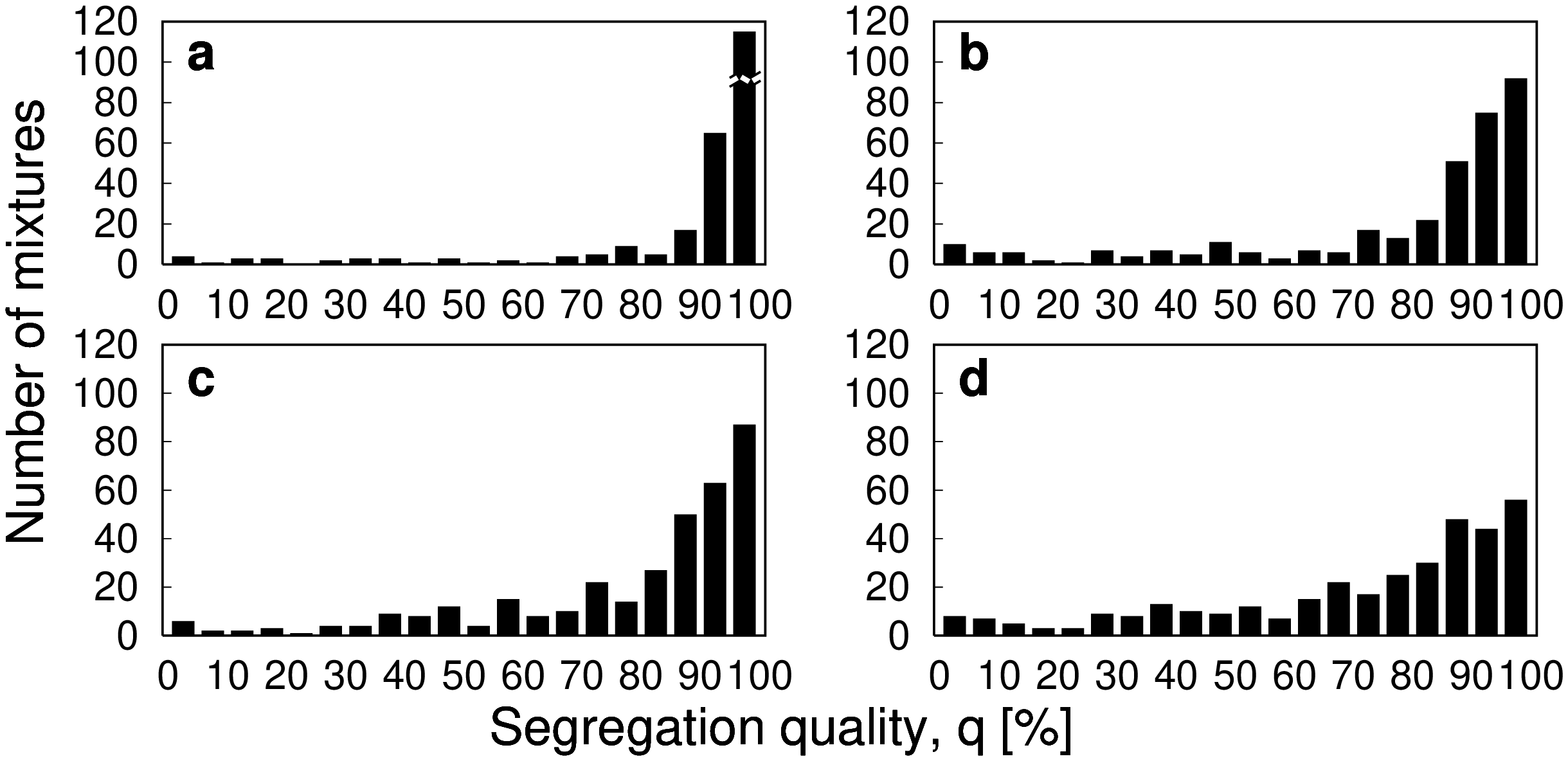,height=1.0\linewidth,angle=270}}}
\newcommand\figsegrall[1]{
%
%
\begin{figure}
#1
\caption{Segregation quality for 351 binary mixtures.
In all cases the segregation parameters are
$\boxwidth = 30 \cm$, $\loadheight = 3 \cm$ and $\loadrate = 0.25 \Hz$.
{\bf a,}
The particles are uniform within the components, and the
rebounding angle noise is $\delta\phi=0$.
The number of mixtures in the last column (quality is better than
95\%) is 219, and the quality is better than 70\% for 320 mixtures.
{\bf b,}
The same as in {\bf a} but the rebounding angle noise is
$\delta\phi=0.3$.
The quality is better than 70\% for 270 mixtures.
{\bf c,}
The same as in {\bf a} but the particles are not uniform within the
components, all of the parameters ($r$, $e$, $\mu$ and $\beta_0$)
are varied by maximum 10\%.
The quality is better than 70\% for 263 mixtures.
{\bf d,}
The rebounding angle noise is $\delta\phi=0.3$ and the particle
parameters are varied by maximum 10\%.
The quality is better than 70\% for 220 mixtures.
}
\label{fig:segrall}
\end{figure} 
}

\nc{\scw[1]}{\hbox to .08\linewidth{\hfill$#1$\hfill}}
\newcommand\tableparameters{
%
%
\begin{table}
\begin{center}
\begin{tabular}{cccccccc}
\scw{r}  & \scw{e} & \scw{\mu} & \scw{w} & \scw{h} & \scw{a} & \scw{f} & \scw{\theta}\\
\mbox{[mm]} &  &  & \mbox{[mm]} & \mbox{[mm]} & & \mbox{[s$^{-1}$]} & \mbox{[rad]}\\
\hline
1 &0.4 & 0.1 & 6  & 6  & 0 & 16 & -0.16\\
1.5 &0.6 & 0.3 & 8  & 8  & 0.1 & 18 & -0.08\\
2 &0.8 & 0.5 & 10  & 10  & 0.2 & 20 & 0.0\\
  &    &     & 12 & 12  & 0.3 & 22 & 0.08\\
  &    &     & 14 & 14 & 0.4 & 24 & 0.16\\
\end{tabular}
\end{center}
\caption{The parameter values used to produce
$3^3=27$ particle types and $5^5=3125$ ratchet types.}
\label{table:parameters}
\end{table}
}

\newcommand\figcaptions{
\figsetup{}
\figdriftdiff{}
\figsegr{}
\figsegrall{}
}
\newcommand\figs{
\newpage
\figsetupfile
\bigskip
\par\noindent
Figure 1\\
Farkas, Z. {\it et al.}\\
``Segregation of granular binary mixtures by a ratchet mechanism''
\newpage
\figdriftdifffile
\bigskip
\par\noindent
Figure 2\\
Farkas, Z. {\it et al.}\\
``Segregation of granular binary mixtures by a ratchet mechanism''
\newpage
\figsegrfile
\bigskip
\par\noindent
Figure 3\\
Farkas, Z. {\it et al.}\\
``Segregation of granular binary mixtures by a ratchet mechanism''
\newpage
\figsegrallfile
\bigskip
\par\noindent
Figure 4\\
Farkas, Z. {\it et al.}\\
``Segregation of granular binary mixtures by a ratchet mechanism''
}


While segregation is often an undesired effect,
sometimes separating the components of a granular mixture
is the ultimate goal.
Since ancient times sieves have been used by humans
to separate small grains from bigger ones.
But nature manages to separate different kinds of grains 
also without sieves.
Many of the segregation processes in granular matter
\cite{jaeger92,mehta94,herrmann98}
have recently been studied in great detail such as
segregation according to
particle size, shape or friction properties
in a shaken box \cite{rosato87,knight93,cooke96},
in a rotating drum \cite{zik94,baumann95,lai97,ktitarev98}
or when poured into a thin box \cite{meakin90,makse97,baxter98}.
However, so far no method has been proposed 
for segregating particles which only differ in hardness.

The setup we use for segregation is as follows:
particles of a binary mixture are falling into a two-dimensional box from
above at a specified place.
The base of the box having an asymmetric sawtooth profile
is shaken harmonically in vertical direction.
The transport properties of homogeneous granular media
in a similar setup were studied earlier \cite{farkas99}
inspired by recent progress in the
theoretical understanding of molecular motors
\cite{julicher97,astumian97}.
In the corresponding models, known as thermal ratchets,
fluctuation-driven transport phenomena can be interpreted in terms of
overdamped Brownian particles moving through a periodic but asymmetric
(typically sawtooth shaped)
potential in the presence of nonequilibrium fluctuating forces (such as
periodic driving or switching between potentials).
When a particle jumps out of the box at either
the right or the left boundary,
it is removed and counted, and finally the segregation quality
is determined.
For details on the geometry see Fig.\ \ref{fig:setup}.
The segregation quality is
$
q = (2\max\{N_{1\leftside} + N_{2\rightside},
          N_{1\rightside} + N_{2\leftside}\} / N - 1) \times 100\%,
$
where $N_{i\leftarrow}$ ($N_{i\rightarrow}$) denotes the number
of particles of the $i$th component ($i=1,2$) leaving the box
on the left (right) side, and $N$ is the number of all particles in
the mixture.
Thus the quality of random segregation, when the particles go to
the left or right side with equal probabilities, is $0\%$.
In this Letter we show results for binary mixtures in which the
components contain equal number of particles,
in other cases different quality definitions may also be appropriate.
\figsetup{\figsetupfile}
Unlike other segregation phenomena, in which segregation is
due to the collective behaviour of the grains, here the
interaction between the base and the individual particles is
dominant,
and the efficiency of segregation usually decreases with
increasing number of particle-particle collisions.
However, setting a sufficiently low load rate $\loadrate$, the
quality can be high, and still, ``parallelizing'' the procedure the
separation capacity can be large (see bellow).
In the simulation an event-driven algorithm \cite{lubachevsky91} is applied
with a hard-sphere collision model \cite{walton}, in which
the sphere-shaped particles have five parameters: mass $m$,
radius $r$, normal restitution coefficient $e$, friction coefficient
$\mu$ and maximum tangential restitution coefficient $\beta_0$.
The particles can rotate around the axis
going through their center and
perpendicular to the plane of the box,
their moment of inertia is
$2/5\, m r^2$ about their center.
Since the mass of the particles does not play a role
in collisions with the base but only in binary collisions,
it is enough to specify that the
particles have the same mass density,
so the mass is cubically proportional to the radius.

If there are only few particles in the box at the same time,
then, as a first approximation, the interaction between
the particles can be neglected.
Therefore, we investigated the motion of one particle
in an infinitely wide box in detail,
and found it to be chaotic \cite{baker90} in most cases.
A similar, but simpler model was also reported to
show chaotic behaviour \cite{duran92}.
Depending on the parameters, it is possible that
the particle follows a periodic trajectory,
travelling with velocity $v = f w b / c$,
where $b$ and $c$ are integer numbers,
meaning that in one period, which lasts for $c$ vibration cycles,
the particle jumps over $b$ teeth.
These periodic trajectories are not interesting for practical applications
for the following reasons:
(1) the transients are usually very long, and therefore may be more
important than the asymptotic periodic trajectory for the segregation
behaviour,
(2) periodic trajectories are not robust against collisions with other
particles and other sources of noise, and, last but not least,
(3) for certain conditions two periodic trajectories with opposite
directions can coexist for the same type of particles.
In this case one cannot predict on which side the
particle tends to leave the box.
We explain below how these periodic trajectories can be
avoided when searching for the ratchet parameters suitable
for segregating a given binary mixture.
In the chaotic regime, however, the time evolution of the
particle's horizontal position can be well described as
drift-diffusion.
The connection between chaotic motion and diffusion has been
investigated extensively recently
\cite{gaspard90,gaspard98,dettmann00}.
A simple explanation for this drift-diffusion behavior can be that
on time scales larger than the typical
time it takes for the particle to jump to another sawtooth,
the kicks of the base can be considered to be independent of each other.
Furthermore, the asymmetry of the ratchet leads to an average velocity
in the left or right direction.
Consequently, the horizontal motion in the chaotic regime
can be described statistically by two parameters:
a drift velocity $v$ and a diffusion coefficient $D$.
An example for the drift-diffusion motion can be seen in Fig.\
\ref{fig:driftdiff}.
\figdriftdiff{\figdriftdifffile}

The observation that the horizontal motion can be well approximated
by drift-diffusion enables us to predict segregation-related
quantities \cite{farkas00}.
First of all,
the probability that a particle jumps out of the box
through the right ($n_\rightarrow$) or the left ($n_\leftarrow$) boundary
is:
$n_\rightarrow(u, L, \ratio) =
\frac{1 - e^{-\ratio L u}}{1 - e^{-L u}}$ and
$n_\leftarrow(u, L, \ratio) = 1 - n_\rightarrow(u, L, \ratio)$,
where we introduce the notation $u = v / D$, since
the probabilities depend on the drift velocity and the diffusion coefficient
only through this combination.
The asymptotic behaviour of these probabilities
in $L$ is exponential with characteristic length $\ratio^{-1}|u|^{-1}$,
which means that if the drift velocity is, say, positive
and the box width is large enough,
most or all of the particles arrive at the right end.
The approximate explanation for this is the following:
the displacement of the average particle position
increases linearly in time,
while the width of the probability distribution is 
proportional only to the square root of time
(although this is exactly true only if the box width is infinite,
for large box widths it is still a good approximation).
Therefore, on length scales larger than $|u|^{-1}$,
the drift dominates over diffusion, so that most likely the particle
leaves the system at the end towards which it drifted.
As a consequence, for the segregation of a binary mixture
the theory suggests that 
if the drift velocities of the particles of the two components
($v_1$ and $v_2$)
have opposite directions, then one can obtain an arbitrarily good
segregation quality by choosing the box wide enough.
For a fixed box width, the best segregation quality is given by
$
q_\idx{opt} = [n_\leftarrow(u_1,L,\ratio_\idx{opt})
+ n_\rightarrow(u_2, L, \ratio_\idx{opt}) - 1]
\times 100\%,
$
where $u_1 < 0 < u_2$, and
$\ratio_\idx{opt} =
1 - \frac{\ln[u_1 (e^{u_2 L} - 1)] /
                  [u_2 (e^{u_1 L} - 1)]}
         {(u_2 - u_1) L}$
gives the optimal place of loading to
obtain the best possible quality.
The asymptotic behaviour of $q_\idx{opt}$ 
as a function of $L$ shows that
it exponentially approaches 100\% with characteristic
length $(|u_1| + |u_2|)\, |u_1 u_2|^{-1}$.
Figure \ref{fig:segr} shows an example for segregation of particles
differing only in friction coefficient.
In this example the particle parameters are:
$r_1 = r_2 = 1.5 \mm$, $e_1 = e_2 = 0.4$, $\mu_1 = 0.1$,
$\mu_2 = 0.3$, $\beta_{01} = \beta_{02} = 0.4$,
the ratchet parameters are:
$w = 8 \mm$, $h = 6 \mm$, $a = 0.4$,
$\theta = -0.08$, $A = 2 \mm$, $f = 20 \Hz$,
and the segregation parameters are: $\ratio = 0.51$,
$\loadheight = 1 \cm$, $\loadrate = 0.5 \Hz$.
The corresponding diffusion parameters are
$\vel_1 = -1.58 \cmps$, $\diffcoeff_1 = 1.36 \cmsps$,
$\vel_2 = 1.86 \cmps$ and $\diffcoeff_2 = 1.83 \cmsps$,
the characteristic length is $1.85 \cm$.
The results are obtained from the segregation of 20,000 particles,
10,000 in both components.
\figsegr{\figsegrfile}

The ratchet segregating a certain binary mixture 
most efficiently is searched in the following way:
for both components the diffusion parameters of the one 
particle motion
($v_1$, $D_1$ and $v_2$, $D_2$)
are measured using many different ratchets.
Then we select those ratchets for which the drift velocities
have opposite directions
(it may happen that no such ratchet is found).
Setting the box width to a reasonable value,
for each of the selected ratchets the best segregation quality is
predicted, and the ratchet with the highest segregation quality is chosen
for segregating the mixture.
However, it is possible that with this ratchet
one or both of the particles have periodic trajectories,
which is undesired for segregation.
We describe here one possible solution to this problem:
any kind of noise can destroy periodic trajectories.
For example, in the computer simulation
the angle of the relative velocity after a particle--base collision
is changed by an amount uniformly chosen from 
an interval $[-\dphi, \dphi]$.
We found that a rebounding angle noise
$\dphi \approx 0.05$ (measured in radian)
is enough to destroy the periodic trajectories,
drastically changing the drift velocity.
Hence, the diffusion parameters for
both particles and each ratchet are also measured
with $\dphi = 0.05$ and $0.1$.
Then, for selecting a ratchet it is not enough that the drift velocities
have opposite direction when $\dphi = 0$,
but their values should not change too much
when the noise level is increased
to $0.05$ and $0.1$
(we allowed a maximum relative change of 50\% and 70\%, respectively).
The best ratchet is chosen from this restricted set by
a rule which takes into account that
(1) the quality should be high also in the noisy case,
(2) $|\vel_1|$ and $|\vel_2|$ should be as large as possible to allow
a high load rate,
(3) $\alpha_\idx{opt}$ should not change much in the presence of noise,
otherwise the segregation quality may decrease much due to interaction
between particles.
This choice of noise is not only good for avoiding periodic
trajectories, but also may serve as a first approximation
for taking into account the deviation
of the particles' shape from a sphere.
Therefore, the ratchets selected by this method are robust
to some extent against deviations in shape and presumably
in other parameters as well.

We checked the practicability of this procedure by
finding appropriate ratchets
to segregate
$\big({27 \atop 2}\big) = 351$
different binary mixtures composed from
27 particle types,
measuring the diffusion parameters for each type
by simulating 3000 seconds of particle trajectories for 3125 ratchets.
The parameters which were varied
are listed in Table \ref{table:parameters},
the maximum tangential restitution coefficient
and the vibration amplitude were fixed
$\beta_0 = 0.4$ and $A = 2$ mm, respectively.
\tableparameters
Then we performed a segregation
simulation for each mixture with the selected ratchet
(for three mixtures no proper ratchet was found in this set).
The results show that in most
cases good segregation quality can be achieved
even if the rebounding angle noise is relatively high or the
particles within the components are not uniform
(see Fig.\ \ref{fig:segrall}).
\figsegrall{\figsegrallfile}

If no ratchet is found, even in an extended set, which segregates a certain
mixture
({\it i.e.,} the drift velocities always have the same sign),
then there is an alternative procedure which we
describe here only briefly.
Unless the particles in the components are exactly the same,
there exists a ratchet for which $u_1$ and $u_2$ are different.
If, {\it e.g.,} $0 < u_1 < u_2$, then choosing an appropriate loading place
near the left end, more particles from component 1
will be collected on the left side than from component 2.
Then the particles collected on the right side are reloaded again and
again. For a given box width and number of reloads one can determine
the optimal load place to obtain the best possible segregation quality,
which tends to 100\% with increasing number of reloads
(results not shown here).

One may think that the method presented
here is not efficient for segregating real granular mixtures,
since only a few grains can be present
in the box at the same time, allowing only a low load rate.
However, it is very easy to ``parallelize'' the procedure:
as the box is essentially two-dimensional, many boxes
can be placed onto a shaking machine, and possibly the walls
between the boxes can be omitted.
A rough estimation shows that
the capacity of such a machine would be comparable to that of the
machines in use today in the industry
for cleaning, {\it e.g.}, cereal grains
(in which case the capacaity is in the order of 1 ton per hour).
We are planning to carry out experiments to check if mixtures containing
particles of the same size differing only in normal restitiution
coefficient or friction coefficient can be segregated by
this method.

We presented a computer simulation study of a method for segregating a
binary granular mixture.
The segregation is performed by a ratchet mechanism, and in contrast to
other segregation schemes, here not the collective behavior of the
particles is dominant but the interaction between the
base and the individual particles.
We found that good segregation quality can be achieved even if the
particles of the two components differ only in friction coefficient or
hardness.

Useful discussions with T. F\"ul\"op and P. Tegzes are acknowledged.
We thank the Computer and Automation Research Institute (Budapest)
for using their PC-cluster.
Z. F. is grateful for financial support by DAAD.
This research has partially been supported by HSF OTKA T033104.\\




\begin{thebibliography}{x}


\bibitem{jaeger92}
H. M. Jaeger, and S. R. Nagel,
Science {\bf 255,} 1523-1531 (1992)

\bibitem{mehta94}
A. Mehta (ed.), {\it Granular Matter: An Interdisciplinary Approach}
(Springer, New York, 1994)

\bibitem{herrmann98}
H. J. Herrmann, J.-P. Hovi, and S. Luding (eds),
{\it Physics of Dry Granular Media}
(Kluwer, Dordrecht, 1998)


\bibitem{rosato87}
A. Rosato, K. J. Strandburg, F. Prinz, and R. H. Swendsen,
\prl {\bf 58,} 1038 (1987)

\bibitem{knight93}
J. B. Knight, H. M. Jaeger, and S. R Nagel,
\prl {\bf 70,} 3728 (1993)

\bibitem{cooke96}
W. Cooke, S. Warr, J. M. Huntley, and R. C. Ball, 
\pre {\bf 53,} 2812 (1996)


\bibitem{zik94}
O. Zik, D. Levine, S. G. Lipson, S. Shtrikman, and J. Stavans, 
\prl {\bf 73,} 644 (1994)

\bibitem{baumann95}
G. Baumann, I. M. Janosi, and D. E. Wolf,
\pre {\bf 51,} 1879 (1995)

\bibitem{lai97}
P.-Y. Lai, L.-C. Jia, and C. K. Chan,
\prl {\bf 79,} 4994 (1997) 

\bibitem{ktitarev98}
D. Ktitarev and D. E. Wolf,
{\it Granular Matter} {\bf 1,} 141 (1998)


\bibitem{meakin90}
P. Meakin,
Physica A {\bf 163,} 733 (1990)

\bibitem{makse97}
H. A. Makse, S. Havlin, P. R. King, and H. E. Stanley,
Nature {\bf 386,} 379 (1997)

\bibitem{baxter98}
J. Baxter, U. Tuzun, D. Heyes, I. Hayati, and P. Fredlund,
Nature {\bf 391,} 136 (1998)

\bibitem{farkas99}
Z. Farkas, P. Tegzes, A. Vukics, and T. Vicsek,
\pre {\bf 60,} 7022 (1999)

\bibitem{julicher97}
F. J\"ulicher, A. Ajdari, and J. Prost,
Rev. Mod. Phys. {\bf 69,} 1269 (1997)

\bibitem{astumian97}
R. D. Astumian,
Science {\bf 276} 917 (1997)

\bibitem{lubachevsky91}
B. D. Lubachevsky,
J.\ Comput.\ Phys. {\bf 94,} 255 (1991)

\bibitem{walton}
O. R. Walton in {\it Particulate Two-Phase Flow}
(ed Roco, O. M.) (Butterworth--Heinemann, Boston, 1992)

\bibitem{baker90}
G. L. Baker and J. P. Gollup,
{\it Chaotic dynamics: an introduction}
(Cambridge University Press, 1990)

\bibitem{duran92}
J. Duran,
Europhys.\ Lett. {\bf 17,} 679 (1992)

\bibitem{gaspard90}
P. Gaspard and G. Nicolis,
\prl {\bf 65,} 1693 (1990)

\bibitem{gaspard98}
P. Gaspard {\it et al.},
Nature {\bf 394,} 865 (1998)

\bibitem{dettmann00}
C. P. Dettmann and E. G. D. Cohen,
{\it nlin.CD/0001062}

\bibitem{risken89}
H. Risken,
{\em The Fokker--Planck Equation}
(Springer, Berlin, 1989)

\bibitem{farkas00}
Z. Farkas and T. F\"ul\"op,
{\it cond-mat/0010358}

\end{thebibliography}
\end{document}